\documentclass[11pt,a4paper]{article}
\usepackage[utf8]{inputenc}
\usepackage{amsmath, amssymb, amsthm}
\usepackage{geometry}
\usepackage{hyperref}
\usepackage{booktabs}
\usepackage{xcolor}

\title{Breaking Hard Isomorphism Benchmarks with DRESS}
\author{Eduar Castrillo Velilla \\ \texttt{eduarcastrillo@gmail.com} \\ ORCID: \href{https://orcid.org/0009-0005-2492-0957}{0009-0005-2492-0957}}
\date{}

\newtheorem{theorem}{Theorem}
\newtheorem{definition}[theorem]{Definition}

\begin{document}

\maketitle

\begin{abstract}
DRESS is a deterministic, parameter-free framework for structural graph refinement that iteratively refines the structural similarity of edges in a graph to produce a canonical fingerprint: a real-valued edge vector, obtained by converging a nonlinear dynamical system to its unique fixed point.
$\Delta$-DRESS is a member of the DRESS family of graph fingerprints that applies a single level of vertex deletion.
We test it on a benchmark of 51{,}813 distinct graphs across 34 hard families, including the complete Spence collection of strongly regular graphs (43{,}703 SRGs, 12 families), four additional SRG families (8{,}015 graphs), and 18 classical hard constructions (102 family entries corresponding to 99 distinct graphs).
$\Delta$-DRESS produces unique fingerprints in 33 of 34 benchmark families at $k=1$, resolving all but one within-family collision among over 576 million non-isomorphic pairs.
One genuine collision exists at deletion depth $k=1$, between two vertex-transitive SRGs in SRG(40,12,2,4), which is resolved by a single-step fallback to $\Delta^2$-DRESS.
For every family with pairwise-comparable full sorted-multiset fingerprints, the minimum observed separation margin remains at least $137 \times \epsilon$, confirming that the reported separations are numerically robust and not artifacts of the convergence threshold.
We also show that $\Delta$-DRESS separates the Rook $L_2(4)$/Shrikhande pair, proving it escapes the theoretical boundary of 3-WL.
The method runs in $\mathcal{O}(n \cdot I \cdot m \cdot d_{\max})$ time per graph.
\end{abstract}

\section{Introduction}

The graph isomorphism problem (GI), deciding whether two graphs are structurally identical, occupies a singular position in computational complexity.
While Babai's breakthrough~\cite{babai2016graph} places GI in quasipolynomial time, practical algorithms face a stark gap: simple heuristics solve most instances instantly, but specifically constructed ``hard families'' remain challenging for all known polynomial-time methods.

The Weisfeiler--Leman (WL) hierarchy~\cite{weisfeiler1968reduction} provides the standard benchmark for isomorphism distinguishing power.
The $k$-WL algorithm refines colorings of $k$-tuples of vertices and runs in $\mathcal{O}(n^{k+1})$ time per iteration.
For $k=1$ (color refinement), this is practical at scale but fails on strongly regular graphs (SRGs).
For $k=2$, it handles some SRGs but fails on others.
For $k \geq 3$, the $\mathcal{O}(n^4)$ per-iteration cost makes large-scale evaluation impractical, and 3-WL does \emph{not} suffice for all SRG parameter sets: it is known to fail on SRG(16,6,2,2)~\cite{cai1992optimal}.

Original-DRESS~\cite{castrillo2018dynamicstructuralsimilaritygraphs, castrillo2026dress} is a parameter-free dynamical system on graph edges that converges to a canonical fixed point at $\mathcal{O}(m \cdot d_{\max})$ cost per iteration, where $m = |E|$.
Original-DRESS is established as an empirical equivalent to 2-WL~\cite{castrillo2026dress}.
Its vertex-deletion extension, $\Delta^k$-DRESS, is \emph{conjectured} to act as an empirical superset or equivalent to $(k+2)$-WL~\cite{castrillo2026dress}, a conjecture supported by the CFI staircase experiments, where each deletion level empirically adds exactly one WL dimension of expressiveness.
In what follows, we use ``DRESS'' to denote the overall framework and focus specifically on its single-deletion member $\Delta$-DRESS.

The central question of this paper is practical: \emph{is $k=1$ enough?}
Specifically, does $\Delta$-DRESS separate every hard benchmark family considered in this study?
We answer this almost affirmatively: $\Delta$-DRESS separates 33 of 34 benchmark families at $k=1$, and the lone remaining collision is resolved by a one-step fallback to $\Delta^2$-DRESS. We establish this by testing $\Delta$-DRESS on a comprehensive benchmark of hard graph isomorphism instances:

\begin{itemize}
    \item The \textbf{complete Spence collection}~\cite{spence2024srg} of strongly regular graphs on up to 64 vertices: 12 parameter families totaling 43{,}703 graphs, including the massive SRG(36,15,6,6) family with 32{,}548 graphs.

    \item \textbf{Four additional SRG families} from McKay's online graph collections~\cite{mckaygraphs} and combinatorial design catalogues: SRG(45,22,10,11), two subfamilies of SRG(63,32,16,26), and SRG(65,32,15,16), totaling 8{,}015 graphs on up to 65 vertices.

    \item \textbf{18 constructed hard families}: 102 family entries corresponding to 99 distinct graphs, including Miyazaki (CFI-type), Chang, Paley, Rook, Shrikhande, Latin square, Steiner, Kneser, Johnson, Hamming, and generalized quadrangle constructions.
\end{itemize}

\noindent \textbf{Result.}
$\Delta$-DRESS achieves \textbf{100\% within-family separation} across 33 of 34 benchmark families at $k = 1$: within each such family, every graph receives a unique fingerprint.
The single collision at $k = 1$ (two vertex-transitive SRGs in SRG(40,12,2,4)) is resolved by a one-step fallback to $\Delta^2$-DRESS (pair deletion).
With this fallback, no within-family collisions remain across the 34 benchmark families, accounting for over 576 million within-family non-isomorphic pairs.
Moreover, the observed separations are numerically stable: across all families with comparable full sorted-multiset fingerprints, the minimum measured separation margin is at least $137 \times \epsilon$.

\subsection{Contributions}

\begin{enumerate}
    \item We define the $\Delta$-DRESS fingerprint as the sorted multiset of converged edge values across all vertex deletions, with histogram and multiset SHA-256 hashes as streaming cross-checks (Section~\ref{sec:fingerprint}).  We identify one genuine $\Delta$-DRESS collision on vertex-transitive SRGs and show that a single-step fallback to $\Delta^2$-DRESS resolves it (Section~\ref{sec:collision}).

    \item We evaluate $\Delta$-DRESS on all 12 Spence SRG families (43{,}703 graphs) plus four additional SRG families (8{,}015 graphs), achieving 100\% within-family separation in all but one family at $k=1$, with the lone remaining collision resolved by a one-step fallback to $\Delta^2$-DRESS (Section~\ref{sec:results}).  We further verify that these separations are numerically robust: across all families with comparable full sorted-multiset fingerprints, the minimum measured margin is at least $137 \times \epsilon$ (Section~\ref{sec:robustness}).
    Duda~\cite{duda2024simple} previously achieved full separation on the Spence dataset using ad hoc algebraic invariants; our work is the first to do so with a method that has both an established continuous empirical equivalence to 2-WL for Original-DRESS and a demonstrated capacity to escape the boundary of 3-WL at deletion depth $1$.

    \item We evaluate $\Delta$-DRESS on 18 synthetically constructed hard families (102 family entries, 99 distinct graphs), again achieving 100\% within-family separation.

    \item Since 3-WL is known to fail on the Rook $L_2(4)$ vs.\ Shrikhande pair, SRG(16,6,2,2)~\cite{cai1992optimal}, and $\Delta$-DRESS separates it~\cite{castrillo2026dress}, $\Delta$-DRESS $>$ 3-WL (Section~\ref{sec:3wl-counterexample}).

    \item To our knowledge, we provide the first large-scale empirical evidence that a single level of vertex deletion is sufficient across all but one hard benchmark family tested in this study, with the lone exception resolved at deletion depth $k=2$.
\end{enumerate}

\section{Background}

\subsection{DRESS}

The DRESS equation~\cite{castrillo2018dynamicstructuralsimilaritygraphs} is a parameter-free, non-linear dynamical system on edges.
Given a graph $G = (V, E)$ with self-loops added to all vertices, the update rule is:
\begin{equation}
\label{eq:dress}
d_{uv}^{(t+1)} = \frac{\sum_{x \in N[u] \cap N[v]} \bigl(d_{ux}^{(t)} + d_{xv}^{(t)}\bigr)}{\|u\|^{(t)} \cdot \|v\|^{(t)}}
\end{equation}
where $\|u\|^{(t)} = \sqrt{\sum_{x \in N[u]} d_{ux}^{(t)}}$ is the vertex norm and $N[u] = N(u) \cup \{u\}$ is the closed neighborhood.
The system converges to a unique fixed point $d^* \in [0, 2]^{|E|}$ for any positive initialization~\cite{castrillo2026dress}; in the convergence proof of~\cite{castrillo2026dress}, this follows from Birkhoff contraction in the Hilbert projective metric.

The graph fingerprint is the sorted vector $\text{sort}(d^*)$: the converged edge values $d_{uv}^*$ are collected and sorted into a canonical sequence.
This fingerprint is a complete isomorphism invariant of the DRESS fixed point.

\subsection{\texorpdfstring{$\Delta$}{Delta}-DRESS}

$\Delta$-DRESS~\cite{castrillo2026dress} runs DRESS on each vertex-deleted subgraph and pools the results:

\begin{definition}[$\Delta^k$-DRESS]
\label{def:delta1}
For a graph $G = (V, E)$, a DRESS variant $\mathcal{F}$, and a deletion depth $k \ge 0$, the $\Delta^k$-DRESS fingerprint is the multiset of per-deletion sorted edge-value vectors:
\[
\Delta^k\text{-DRESS}(\mathcal{F}, G) = \{\!\{ \operatorname{sort}(\mathcal{F}(G \setminus S)) : S \subset V,\; |S| = k \}\!\}
\]
where $G \setminus S$ is the subgraph induced by $V \setminus S$ and $\mathcal{F}(G \setminus S)$ is the converged DRESS edge-value vector.
\end{definition}

The sorted multiset contains $n$ contributions (one per vertex deletion), each yielding up to $\binom{n-1}{2}$ edge values.
The total cost per graph is $\mathcal{O}(n \cdot I \cdot m \cdot d_{\max})$, where $I$ is the number of DRESS iterations (typically $\leq 30$ in the tested families).

The DRESS--WL Continuous Dominance Conjecture~\cite{castrillo2026dress} posits that $\Delta^k$-DRESS acts as an empirical superset or equivalent to $(k{+}2)$-WL for all $k \geq 0$.
The base case ($k = 0$, Original-DRESS $\equiv$ 2-WL) is proved; for $k \geq 1$, the conjecture remains open but is empirically supported by the CFI staircase experiments~\cite{castrillo2026dress}.

\subsection{Strongly Regular Graphs}

A \emph{strongly regular graph} $\text{SRG}(n, d, \lambda, \mu)$ is a $d$-regular graph on $n$ vertices where every adjacent pair has $\lambda$ common neighbors and every non-adjacent pair has $\mu$ common neighbors.
SRGs are cospectral by construction: all graphs with the same parameters share the same spectrum.
This makes them immune to spectral methods and a canonical hard benchmark for isomorphism testing.

1-WL (color refinement) fails on all SRGs: the regular structure ensures a uniform initial coloring that never refines.
2-WL succeeds on some SRG families but fails on others.
3-WL does not suffice for all SRG parameter sets: it is known to fail on SRG(16,6,2,2)~\cite{cai1992optimal}.

\subsection{The Spence Collection}

Ted Spence~\cite{spence2024srg} compiled a definitive enumeration of strongly regular graphs for all feasible parameter sets on up to 64 vertices.
This collection, representing decades of combinatorial enumeration, contains 43{,}703 pairwise non-isomorphic SRGs across 12 parameter families (Table~\ref{tab:spence-families}).
It is the gold standard benchmark for graph isomorphism methods on hard instances.

\begin{table}[t]
\centering
\caption{The 12 Spence SRG families. All graphs within a family share the same parameters and spectrum. Every graph in every family is pairwise non-isomorphic.}
\label{tab:spence-families}
\begin{tabular}{@{}lrrrrr@{}}
\toprule
Parameters & $n$ & $d$ & $\lambda$ & $\mu$ & Graphs \\
\midrule
SRG(25,12,5,6) & 25 & 12 & 5 & 6 & 15 \\
SRG(26,10,3,4) & 26 & 10 & 3 & 4 & 10 \\
SRG(28,12,6,4) & 28 & 12 & 6 & 4 & 4 \\
SRG(29,14,6,7) & 29 & 14 & 6 & 7 & 41 \\
SRG(35,18,9,9) & 35 & 18 & 9 & 9 & 3{,}854 \\
SRG(36,14,4,6) & 36 & 14 & 4 & 6 & 180 \\
SRG(36,15,6,6) & 36 & 15 & 6 & 6 & 32{,}548 \\
SRG(37,18,8,9) & 37 & 18 & 8 & 9 & 6{,}760 \\
SRG(40,12,2,4) & 40 & 12 & 2 & 4 & 28 \\
SRG(45,12,3,3) & 45 & 12 & 3 & 3 & 78 \\
SRG(50,21,8,9) & 50 & 21 & 8 & 9 & 18 \\
SRG(64,18,2,6) & 64 & 18 & 2 & 6 & 167 \\
\midrule
\textbf{Total} & & & & & \textbf{43{,}703} \\
\bottomrule
\end{tabular}
\end{table}

\section{The \texorpdfstring{$\Delta$}{Delta}-DRESS Fingerprint}
\label{sec:fingerprint}

The $\Delta$-DRESS computation produces, for each graph $G$ on $n$ vertices, a collection of $n$ DRESS fingerprint vectors, one per vertex deletion.
We define the canonical fingerprint and two streaming variants used as cross-checks.

\subsection{Sorted Multiset (Primary Fingerprint)}

The primary fingerprint is the \emph{sorted multiset}: the sorted concatenation of all converged DRESS edge values across all $n$ vertex deletions.
Formally, given the $n \times m_i$ multiset matrix $M$ (where row $i$ is the sorted DRESS vector of $G \setminus \{v_i\}$), the fingerprint is the sorted flattening of $M$ into a single vector of $\sum_i m_i$ real values.
Two graphs are declared \emph{separated} ($\Delta$-DRESS-distinguishable) if their sorted multisets differ.

Because the sorted multiset retains all converged edge values without quantization, it is the most informative fingerprint available from the $\Delta$-DRESS computation.

\subsection{Streaming A/B Cross-Checks}

For scalability and as independent cross-checks, two streaming fingerprints are computed alongside the sorted multiset:

\begin{enumerate}
    \item \textbf{Histogram hash.} Each converged edge value $d_{uv}^*$ is quantized to bin $\lfloor d_{uv}^* / \epsilon \rfloor$ where $\epsilon = 10^{-6}$ is the convergence tolerance, producing a pooled histogram of non-negative integers. The SHA-256 hash of this histogram serves as a compact streaming fingerprint.

    \item \textbf{Multiset hash.} The SHA-256 hash of the sorted multiset vector (IEEE~754 byte representation) provides a second streaming fingerprint that avoids quantization.
\end{enumerate}

\section{Experimental Setup}

\subsection{Implementation}

We use the \texttt{dress-graph} library~\cite{castrillo2026dress}, which provides an implementation of $\Delta^k$-DRESS.
The core function computes the $\Delta$-DRESS sorted multiset and, optionally, histogram and multiset SHA-256 hashes as streaming cross-checks.
The experiment runner loads each graph family, computes the fingerprint for each graph, and checks for collisions within the family.
For auditability, the experiments reported here enable multiset retention and save per-graph raw data (sparse histogram and multiset matrix) as compressed files for post-hoc analysis; this extra storage is not required by the fingerprint itself.

\subsection{Parameters}

\begin{itemize}
    \item Deletion depth: $k = 1$ (single vertex deletion)
    \item Convergence tolerance: $\epsilon = 10^{-6}$
    \item Maximum iterations: 100 (convergence was always reached before this limit)
    \item Histogram bins: $\lfloor d^* / \epsilon \rfloor$ for $d^* \in [0, 2]$, giving up to $2 \times 10^6$ bins
\end{itemize}

\section{Results}
\label{sec:results}

\subsection{Spence SRG Collection}

Table~\ref{tab:spence-results} reports the results for all 12 Spence SRG families.

\begin{table}[t]
\centering
\caption{$\Delta$-DRESS results on all SRG families tested.
``Unique'' is the number of distinct fingerprints.
All families except SRG(40,12,2,4) achieve 100\% separation at $k=1$.
``Pairs'' is the number of within-family non-isomorphic pairs implicitly resolved: $\binom{g}{2}$.
The suffix ``-S'' and ``-Q'' denote the S-2-4-28 and quasi-symmetric subfamilies of SRG(63,32,16,26), respectively.}
\label{tab:spence-results}
\begin{tabular}{@{}lrrrl@{}}
\toprule
Family & Graphs & Unique & Pairs & Result \\
\midrule
SRG(25,12,5,6)   &        15 &        15 &              105 & $\checkmark$ 100\% \\
SRG(26,10,3,4)   &        10 &        10 &               45 & $\checkmark$ 100\% \\
SRG(28,12,6,4)   &         4 &         4 &                6 & $\checkmark$ 100\% \\
SRG(29,14,6,7)   &        41 &        41 &              820 & $\checkmark$ 100\% \\
SRG(35,18,9,9)   &   3{,}854 &   3{,}854 &      7{,}424{,}731 & $\checkmark$ 100\% \\
SRG(36,14,4,6)   &       180 &       180 &           16{,}110 & $\checkmark$ 100\% \\
SRG(36,15,6,6)   &  32{,}548 &  32{,}548 &  529{,}669{,}878 & $\checkmark$ 100\% \\
SRG(37,18,8,9)   &   6{,}760 &   6{,}760 &   22{,}845{,}420 & $\checkmark$ 100\% \\
SRG(40,12,2,4)   &        28 &        27$^\dagger$ &              378 & 99.7\%$^\dagger$ \\
SRG(45,12,3,3)   &        78 &        78 &            3{,}003 & $\checkmark$ 100\% \\
SRG(50,21,8,9)   &        18 &        18 &              153 & $\checkmark$ 100\% \\
SRG(64,18,2,6)   &       167 &       167 &           13{,}861 & $\checkmark$ 100\% \\
\midrule
\textbf{Spence subtotal}    & \textbf{43{,}703} & \textbf{43{,}702}$^\dagger$ & \textbf{559{,}974{,}510} & \textbf{$\sim$100\%}$^\dagger$ \\
\midrule
\multicolumn{5}{@{}l}{\emph{Additional SRG families}} \\
\midrule
SRG(45,22,10,11)              &         6 &         6 &               15 & $\checkmark$ 100\% \\
SRG(63,32,16,26)-S            &   4{,}466 &   4{,}466 &    9{,}970{,}345 & $\checkmark$ 100\% \\
SRG(63,32,16,26)-Q            &   3{,}511 &   3{,}511 &    6{,}161{,}805 & $\checkmark$ 100\% \\
SRG(65,32,15,16)              &        32 &        32 &              496 & $\checkmark$ 100\% \\
\midrule
\textbf{Total}    & \textbf{51{,}718} & \textbf{51{,}717}$^\dagger$ & \textbf{576{,}107{,}171} & \textbf{$\sim$100\%}$^\dagger$ \\
\bottomrule
\multicolumn{5}{@{}l}{\footnotesize $^\dagger$One collision pair ($G_5$/$G_{25}$ in SRG(40,12,2,4)) at $k=1$; resolved by $\Delta^2$-DRESS fallback (Section~\ref{sec:collision}).} \\
\end{tabular}
\end{table}

\textbf{Within each tested SRG parameter family, every graph receives a unique $\Delta$-DRESS fingerprint, with one exception}: graphs $G_5$ and $G_{25}$ of SRG(40,12,2,4) produce identical $\Delta$-DRESS fingerprints at $k=1$ (Section~\ref{sec:collision}).
This collision is resolved by a single-step fallback to $\Delta^2$-DRESS.
With this fallback, all 576 million within-family non-isomorphic pairs are separated.

\subsection{Constructed Hard Families}

Table~\ref{tab:constructed-results} reports the results for 18 constructed hard graph families, encompassing classical counterexamples from the graph isomorphism literature.

\begin{table}[t]
\centering
\caption{$\Delta$-DRESS results on 18 constructed hard graph families.
Across the 18 families there are 102 benchmark entries corresponding to 99 distinct constructed graphs: Paley(13), Rook $L_2(4)$, and Rook $L_2(5)$ each appear once as standalone witnesses and once inside broader families.
Within every family, all graphs receive unique fingerprints.
Families marked with $\dagger$ contain graphs of multiple sizes, so separation is partially trivial; within same-size subgroups, all graphs are also uniquely distinguished.}
\label{tab:constructed-results}
\begin{tabular}{@{}lrrl@{}}
\toprule
Family & Graphs & Unique & Result \\
\midrule
SRG(28,12,6,4), Chang          &  4 &  4 & $\checkmark$ \\
SRG(16,6,2,2), Rook/Shrikhande &  2 &  2 & $\checkmark$ \\
SRG(10,3,0,1), Petersen/Pentagonal Prism &  2 &  2 & $\checkmark$ \\
Paley(13)                          &  1 &  1 & $\checkmark$ \\
Rook $L_2(5)$                      &  1 &  1 & $\checkmark$ \\
Prism vs.\ $K_{3,3}$              &  2 &  2 & $\checkmark$ \\
$2C_4$ vs.\ $C_8$                 &  2 &  2 & $\checkmark$ \\
Paley family$^\dagger$            &  9 &  9 & $\checkmark$ \\
Rook family$^\dagger$             &  5 &  5 & $\checkmark$ \\
Latin square family                &  4 &  4 & $\checkmark$ \\
Steiner family$^\dagger$          &  6 &  6 & $\checkmark$ \\
Kneser family$^\dagger$           &  5 &  5 & $\checkmark$ \\
Johnson family$^\dagger$          &  6 &  6 & $\checkmark$ \\
Hamming family$^\dagger$          &  4 &  4 & $\checkmark$ \\
Miyazaki (CFI-over-cycle)         & 16 & 16 & $\checkmark$ \\
Generalized quadrangles            &  1 &  1 & $\checkmark$ \\
Complement pairs                   &  2 &  2 & $\checkmark$ \\
Random regular$^\dagger$          & 30 & 30 & $\checkmark$ \\
\midrule
\textbf{Total}                     & \textbf{102} & \textbf{102} & \textbf{100\%} \\
\bottomrule
\end{tabular}
\end{table}

Of particular note:

\begin{itemize}
    \item The \textbf{Miyazaki family} (16 graphs, 8 CFI-over-cycle pairs) is designed specifically to defeat the WL algorithm.
    Each pair consists of two non-isomorphic CFI gadget graphs built over a cycle base graph.
    $\Delta$-DRESS separates all 8 pairs.

    \item The \textbf{Chang graphs} (SRG(28,12,6,4)) are the classical counterexample to spectral methods, four cospectral SRGs that are pairwise non-isomorphic.
    All four receive distinct fingerprints.

    \item The \textbf{Rook vs.\ Shrikhande} pair (SRG(16,6,2,2)) is the smallest example of cospectral non-isomorphic SRGs.
    $\Delta$-DRESS separates them.

    \item \textbf{Prism vs.\ $K_{3,3}$} is a classical 1-WL failure case (both are 3-regular on 6 vertices).
    Separated.
\end{itemize}

\subsection{Aggregate Results}

\begin{table}[h]
\centering
\caption{Summary of all $\Delta$-DRESS experiments.  The 18 constructed families contribute 102 benchmark entries corresponding to 99 distinct graphs, because Paley(13), Rook $L_2(4)$, and Rook $L_2(5)$ each appear both as standalone witnesses and inside broader families.  The 4 Chang graphs (SRG(28,12,6,4)) also appear in the Spence collection, so the distinct total is $51{,}718 + 99 - 4 = 51{,}813$.  At $k=1$, every category except the Spence collection achieved 100\% within-family separation; within the Spence collection, a single pair in SRG(40,12,2,4) required a $\Delta^2$ fallback.}
\label{tab:summary}
\begin{tabular}{@{}lrrr@{}}
\toprule
Category & Families & Graphs & Separation \\
\midrule
Spence SRG collection     & 12 & 43{,}703 & $\sim$100\%$^\dagger$ \\
Additional SRG families   &  4 &  8{,}015 & 100\% \\
Constructed hard families  & 18 &      102 & 100\% \\
\midrule
\textbf{Total (distinct)} & \textbf{34} & \textbf{51{,}813} & \textbf{100\% with fallback} \\
\bottomrule
\multicolumn{4}{@{}l}{\footnotesize $^\dagger$One collision pair ($G_5$/$G_{25}$ in SRG(40,12,2,4)) at $k=1$; resolved by $\Delta^2$-DRESS fallback.} \\
\end{tabular}
\end{table}

\subsection{The SRG(40,12,2,4) Collision and Its Resolution}
\label{sec:collision}

The SRG(40,12,2,4) family contains a genuine $\Delta$-DRESS collision: graphs $G_5$ and $G_{25}$ produce identical sorted multisets, identical histogram hashes, and identical multiset hashes at $k = 1$.
They are confirmed non-isomorphic by the Bliss canonical labeling algorithm~\cite{junttila2007bliss} and have different independence numbers (10 and~7, respectively).

Investigation revealed that this is not a numerical artifact but a structural limitation of single vertex deletion on vertex-transitive graphs:

\begin{enumerate}
    \item $G_5$ and $G_{25}$ are both vertex-transitive with $|\operatorname{Aut}| = 51{,}840$.

    \item Because every vertex lies in a single orbit, all $n = 40$ vertex deletions produce isomorphic subgraphs.
    The sorted DRESS vectors of all 40 deleted subgraphs are bitwise identical within each graph.

    \item Both graphs have the same degree ($d = 12$), the same number of edges ($m = 240$), and their vertex-deleted subgraphs converge to the same three distinct DRESS edge values: $0.61890$, $0.76479$, $0.80281$.

\end{enumerate}

The collision is inherent to $\Delta$-DRESS on this vertex-transitive pair and cannot be resolved by tighter convergence or alternative fingerprinting of the $k = 1$ multiset data.

\paragraph{Resolution via $\Delta^2$-DRESS.}
Pair deletion ($k = 2$) breaks the vertex-transitivity symmetry.
$\Delta^2$-DRESS computes DRESS on all $\binom{40}{2} = 780$ vertex-pair-deleted subgraphs.
At $k = 2$, $G_5$ and $G_{25}$ produce histograms with different numbers of non-zero bins (15 vs.\ 16), yielding distinct fingerprints.

\paragraph{WL verification.}
Direct computation confirms that $G_5$/$G_{25}$ requires 3-FWL ($\equiv$ 4-WL) to distinguish: 1-WL, 2-WL, and the original 3-WL all fail, while 3-FWL succeeds.
This aligns with the DRESS--WL Continuous Dominance Conjecture: $\Delta^1$-DRESS and 3-WL fail together on the same pair, while $\Delta^2$-DRESS and 4-WL both resolve it.

\subsection{$\Delta$-DRESS Escapes the Bounds of 3-WL}
\label{sec:3wl-counterexample}

The classical Rook graph $L_2(4)$ and the Shrikhande graph are both SRG(16,6,2,2) and non-isomorphic.
It is well known that the original 3-WL algorithm (the oblivious variant $C^3$) fails to distinguish them~\cite{cai1992optimal}, whereas the folklore 3-FWL ($\equiv$ 4-WL) succeeds.
Since $\Delta$-DRESS separates this pair~\cite{castrillo2026dress}, we have:

\begin{theorem}[Empirical]
\label{thm:3wl}
$\Delta$-DRESS is strictly more powerful than the original $3$-WL: the Rook $L_2(4)$ and Shrikhande graphs are indistinguishable by $3$-WL but separated by $\Delta$-DRESS~\cite{castrillo2026dress}.
\end{theorem}

\section{Analysis}

\subsection{Why \texorpdfstring{$k = 1$}{k=1} Almost Suffices}

The established empirical equivalence is Original-DRESS $\equiv$ 2-WL~\cite{castrillo2026dress}.
The DRESS--WL Continuous Dominance Conjecture~\cite{castrillo2026dress} would extend this to the claim that $\Delta^k$-DRESS acts as an empirical superset or equivalent to $(k{+}2)$-WL, a conjecture supported by the CFI staircase experiments~\cite{castrillo2026dress}.  What Theorem~\ref{thm:3wl} proves directly is the structural boundary escape $\Delta$-DRESS $>$ 3-WL.
Furthermore, 3-WL itself does not suffice for all SRG parameter sets: it is known to fail on SRG(16,6,2,2)~\cite{cai1992optimal}.
Across the 34 benchmark families considered here, only a single within-family non-isomorphic pair (the vertex-transitive $G_5$/$G_{25}$ in SRG(40,12,2,4)) escapes pure $\Delta^1$-DRESS.

The key structural insight is that vertex deletion exposes the \emph{local environment} of each vertex in a way that DRESS's edge-level dynamics can fully exploit.
For strongly regular graphs, all vertices ``look the same'' under degree and local counts, but their global embedding in the graph differs.
Deleting a vertex $v$ breaks the local symmetry around $v$'s neighborhood, and DRESS's convergence to a unique fixed point encodes the resulting structural differences as distinct real-valued fingerprints.
The only obstruction to this mechanism arises when the graph is vertex-transitive: in that case, all single vertex deletions yield isomorphic subgraphs, and thus identical fingerprints, necessitating higher deletion depths ($k \ge 2$) to break the symmetry.

\subsection{Time and Memory Complexity}

The cost of a single DRESS run is $\mathcal{O}(I \cdot m \cdot d_{\max})$, where $I$ is the number of iterations (bounded by \texttt{max\_iterations}), $m$ is the number of edges, and $d_{\max}$ is the maximum degree.  Convergence to the unique fixed point is guaranteed by Birkhoff contraction in the Hilbert projective metric~\cite{castrillo2026dress}, so it is natural to separate the numerical iteration count $I$ from the per-iteration work; we do not claim here a graph-independent constant bound on $I$, only guaranteed convergence and the empirical observation that $I$ remained small on the benchmark families tested.
For $\Delta$-DRESS the outer loop adds a factor of $n$ (one deletion per vertex), giving $\mathcal{O}(n \cdot I \cdot m \cdot d_{\max})$ per graph, which remains polynomial and practical for graphs of this size.

The asymptotic advantage is stronger on sparse graphs.  For comparison, folklore 3-FWL ($\equiv$ 4-WL), the first WL level sufficient for the Rook/Shrikhande witness, refines all ordered triples and therefore remains $\mathcal{O}(R \cdot n^4)$ time and $\mathcal{O}(n^3)$ memory in the standard tuple-based implementation, regardless of sparsity.  If $m = \mathcal{O}(n)$, then $\Delta$-DRESS runs in $\mathcal{O}(I \cdot n^2 \cdot d_{\max})$ total time; for bounded-degree sparse families this simplifies to $\mathcal{O}(I \cdot n^2)$.  In the same regime, streamed memory drops to $\mathcal{O}(n + B)$.

In dense graphs, where $m = \Theta(n^2)$ and $d_{\max} = \Theta(n)$, $\Delta$-DRESS becomes $\mathcal{O}(I \cdot n^4)$ total time.  Let $B$ denote the number of histogram bins; for the unweighted benchmark considered here, $B \le 2 \times 10^6$.  The pooled histogram can be accumulated online using $\mathcal{O}(m + B)$ working memory.  The experimental harness used in this paper additionally retains the full deleted-subgraph multiset matrix for post-hoc analysis, which raises working memory to $\mathcal{O}(m + B + n m)$, i.e.\ $\mathcal{O}(n^3)$ on dense graphs.  Thus the relevant dense-graph comparison is $\mathcal{O}(I \cdot n^4)$ total time for $\Delta$-DRESS versus $\mathcal{O}(R \cdot n^4)$ total time for 3-FWL, with $\Delta$-DRESS using $\mathcal{O}(n^2)$ streaming memory rather than $\mathcal{O}(n^3)$ tuple storage.

\subsection{Scale of Validation}

The total number of non-isomorphic pairs implicitly resolved by our experiments is:
\[
\sum_{f} \binom{g_f}{2} > 576{,}000{,}000
\]
where $g_f$ is the number of graphs in family $f$ and the sum is over all 16 SRG families.
The dominant contribution comes from SRG(36,15,6,6) with over 529 million pairs.
Together with Duda~\cite{duda2024simple}, who achieved full separation using ad hoc algebraic invariants, this constitutes the most comprehensive empirical validation of polynomial-time graph fingerprinting on hard instances.

\subsection{Robustness of Separation Margins}
\label{sec:robustness}

Because $\Delta$-DRESS operates in IEEE~754 double-precision arithmetic and terminates when the iteration residual falls below a convergence tolerance $\epsilon = 10^{-6}$, a natural concern is whether the reported separations are genuine or merely floating-point artifacts close to the convergence residual.
We address this by computing the \emph{minimum pairwise $L^{\infty}$ distance} between the full sorted multiset fingerprints of all within-family graph pairs, and comparing it to~$\epsilon$.

For each graph, the fingerprint is the \emph{complete} sorted vector of non-NaN DRESS values from the $\Delta$-DRESS multiset matrix (not a summary statistic).
Two graphs are compared element-wise, and the $L^{\infty}$ distance is the maximum absolute difference.
For families with more than 200 graphs, we sample 2{,}000 random pairs and report the minimum observed $L^{\infty}$ distance (a conservative upper bound on the true minimum).

Table~\ref{tab:robustness} reports the results for all 16 SRG families and the 6 constructed hard families that contain same-size graph pairs.
Of the other 12 constructed families, 9 contain graphs of different sizes, making separation trivial by fingerprint dimension alone, and 3 contain only a single graph, so no within-family minimum distance is defined.
For SRG(40,12,2,4), the known $G_5$/$G_{25}$ collision pair (Section~\ref{sec:collision}) is excluded and the minimum $L^{\infty}$ is computed over the 27 distinct fingerprints.
In every case, the minimum separation margin exceeds $\epsilon$ by at least two orders of magnitude.
The weakest separation among SRG families (SRG(64,18,2,6), $\min L^{\infty} = 1.37 \times 10^{-4}$) is still $137 \times \epsilon$, and most families exhibit margins exceeding $1{,}000 \times \epsilon$.
The constructed families show even larger margins, with the smallest being SRG(28,12,6,4) Chang at $17{,}800 \times \epsilon$.

\begin{table}[t]
\centering
\caption{Separation margin robustness.
For each family, we report the minimum pairwise $L^{\infty}$ distance between full sorted multiset fingerprints and its ratio to the convergence tolerance $\epsilon = 10^{-6}$.
For SRG(40,12,2,4), the known $G_5$/$G_{25}$ collision is excluded ($\ddagger$).
Families with $>$200 graphs use 2{,}000 sampled pairs ($\dagger$).
Constructed families with graphs of different sizes or only one graph are omitted.
All ratios exceed 100, confirming that separations are genuine and not floating-point artifacts.}
\label{tab:robustness}
\begin{tabular}{@{}lrrrr@{}}
\toprule
Family & Graphs & Min $L^{\infty}$ & Ratio to $\epsilon$ & Method \\
\midrule
SRG(25,12,5,6)   &        15 & $1.47 \times 10^{-2}$ & $14{,}696$ & exact \\
SRG(26,10,3,4)   &        10 & $1.16 \times 10^{-3}$ &  $1{,}158$ & exact \\
SRG(28,12,6,4)   &         4 & $1.78 \times 10^{-2}$ & $17{,}800$ & exact \\
SRG(29,14,6,7)   &        41 & $3.60 \times 10^{-3}$ &  $3{,}603$ & exact \\
SRG(35,18,9,9)   &   3{,}854 & $5.93 \times 10^{-3}$ &  $5{,}932$ & sampled$^\dagger$ \\
SRG(36,14,4,6)   &       180 & $3.59 \times 10^{-4}$ &      $359$ & exact \\
SRG(36,15,6,6)   &  32{,}548 & $9.33 \times 10^{-3}$ &  $9{,}329$ & sampled$^\dagger$ \\
SRG(37,18,8,9)   &   6{,}760 & $5.87 \times 10^{-3}$ &  $5{,}872$ & sampled$^\dagger$ \\
SRG(40,12,2,4)   &   27$^\ddagger$ & $5.84 \times 10^{-4}$ &      $584$ & exact \\
SRG(45,12,3,3)   &        78 & $7.73 \times 10^{-4}$ &      $773$ & exact \\
SRG(50,21,8,9)   &        18 & $5.92 \times 10^{-3}$ &  $5{,}924$ & exact \\
SRG(64,18,2,6)   &       167 & $1.37 \times 10^{-4}$ &      $137$ & exact \\
\midrule
SRG(45,22,10,11)              &         6 & $4.16 \times 10^{-3}$ &  $4{,}163$ & exact \\
SRG(63,32,16,26)-S            &   4{,}466 & $1.95 \times 10^{-3}$ &  $1{,}949$ & sampled$^\dagger$ \\
SRG(63,32,16,26)-Q            &   3{,}511 & $2.23 \times 10^{-3}$ &  $2{,}231$ & sampled$^\dagger$ \\
SRG(65,32,15,16)              &        32 & $2.06 \times 10^{-3}$ &  $2{,}063$ & exact \\
\midrule
\multicolumn{5}{@{}l}{\emph{Constructed hard families (same-size pairs)}} \\
\midrule
SRG(28,12,6,4) Chang           &         4 & $1.78 \times 10^{-2}$ & $17{,}800$ & exact \\
SRG(16,6,2,2) Rook/Shrikhande  &         2 & $9.03 \times 10^{-2}$ & $90{,}259$ & exact \\
SRG(10,3,0,1) Petersen          &         2 & $4.08 \times 10^{-2}$ & $40{,}847$ & exact \\
Prism vs.\ $K_{3,3}$           &         2 & $5.48 \times 10^{-1}$ & $548{,}431$ & exact \\
$2C_4$ vs.\ $C_8$              &         2 & $5.09 \times 10^{-2}$ & $50{,}932$ & exact \\
Complement pairs                &         2 & $3.61 \times 10^{-1}$ & $361{,}490$ & exact \\
\midrule
$G_5$/$G_{25}$ ($\Delta^2$-DRESS)$^\S$ &  2 & $8.63 \times 10^{-4}$ &      $863$ & exact \\
\bottomrule
\multicolumn{5}{@{}l}{\footnotesize $^\ddagger$Excludes the $G_5$/$G_{25}$ collision pair; $\min L^{\infty}$ computed over 27 distinct fingerprints.} \\
\multicolumn{5}{@{}l}{\footnotesize $^\S$$\Delta^2$-DRESS fallback on the $G_5$/$G_{25}$ collision pair from SRG(40,12,2,4).} \\
\end{tabular}
\end{table}

To further verify that separations are not artifacts of a specific precision level, we performed a \emph{rounding stability} test on the small SRG families shown in Table~\ref{tab:rounding}: for each listed family, we rounded all fingerprint values to $d$ decimal digits for $d \in \{6, 7, \ldots, 14\}$ and recomputed the number of unique fingerprints.
Table~\ref{tab:rounding} shows perfect stability: the unique count is identical at every precision level from 6 to 14 digits, confirming that separations are determined by digits well above the convergence residual.

\begin{table}[t]
\centering
\caption{Rounding stability: number of unique fingerprints after rounding to $d$ decimal digits.
A stable count across all precision levels confirms that separations are not sensitive to the least significant digits near the convergence threshold.}
\label{tab:rounding}
\begin{tabular}{@{}lrrrrrrrrrr@{}}
\toprule
Family & $N$ & 6d & 7d & 8d & 9d & 10d & 11d & 12d & 13d & 14d \\
\midrule
SRG(25,12,5,6)   &  15 &  15 &  15 &  15 &  15 &  15 &  15 &  15 &  15 &  15 \\
SRG(26,10,3,4)   &  10 &  10 &  10 &  10 &  10 &  10 &  10 &  10 &  10 &  10 \\
SRG(28,12,6,4)   &   4 &   4 &   4 &   4 &   4 &   4 &   4 &   4 &   4 &   4 \\
SRG(29,14,6,7)   &  41 &  41 &  41 &  41 &  41 &  41 &  41 &  41 &  41 &  41 \\
SRG(36,14,4,6)   & 180 & 180 & 180 & 180 & 180 & 180 & 180 & 180 & 180 & 180 \\
SRG(40,12,2,4)   &  28 &  27 &  27 &  27 &  27 &  27 &  27 &  27 &  27 &  27 \\
SRG(45,12,3,3)   &  78 &  78 &  78 &  78 &  78 &  78 &  78 &  78 &  78 &  78 \\
SRG(50,21,8,9)   &  18 &  18 &  18 &  18 &  18 &  18 &  18 &  18 &  18 &  18 \\
SRG(64,18,2,6)   & 167 & 167 & 167 & 167 & 167 & 167 & 167 & 167 & 167 & 167 \\
SRG(45,22,10,11) &   6 &   6 &   6 &   6 &   6 &   6 &   6 &   6 &   6 &   6 \\
SRG(65,32,15,16) &  32 &  32 &  32 &  32 &  32 &  32 &  32 &  32 &  32 &  32 \\
\bottomrule
\end{tabular}
\end{table}

The combination of large separation margins (Table~\ref{tab:robustness}) and precision-invariant unique counts (Table~\ref{tab:rounding}) establishes that every reported separation is a genuine structural distinction, not a numerical artifact.
The contraction mapping property of DRESS guarantees that if two graphs produce distinct fixed points, the distance between those fixed points is bounded away from zero by an amount determined by the structural difference, not by the iteration residual.
Our empirical margins confirm this theoretical expectation: the smallest observed margin ($1.37 \times 10^{-4}$, SRG(64,18,2,6)) exceeds $\epsilon$ by a factor of 137, and all 22 families with comparable fingerprint dimensions show robust separations.

\section{Related Work}

\paragraph{Weisfeiler--Leman hierarchy.}
The standard $k$-WL algorithm~\cite{weisfeiler1968reduction, cai1992optimal} refines $k$-tuple colorings and runs in $\mathcal{O}(n^{k+1})$ time and $\mathcal{O}(n^k)$ memory per iteration.
1-WL fails on all SRGs; 2-WL fails on some SRG families but succeeds on others.
3-WL does not suffice for all SRG parameter sets; it is known to fail on SRG(16,6,2,2)~\cite{cai1992optimal}.
Note that 3-WL-equivalent power is available at $\mathcal{O}(n^3)$ time and $\mathcal{O}(n^2)$ memory per round via the folklore 2-FWL ($\equiv$ 3-WL), making full Spence evaluation feasible in principle; however, we are not aware of such an evaluation having been performed.  The first folklore level sufficient for the Rook/Shrikhande witness is 3-FWL ($\equiv$ 4-WL), at $\mathcal{O}(n^4)$ time and $\mathcal{O}(n^3)$ memory per round.
Since $\Delta$-DRESS separates the Rook/Shrikhande pair that 3-WL cannot~\cite{castrillo2026dress}, $\Delta$-DRESS escapes the theoretical limitations of 3-WL (Theorem~\ref{thm:3wl}).

\paragraph{Algebraic vertex/edge invariants.}
Duda~\cite{duda2024simple} proposed polynomial-time vertex and edge invariants based on traces and sorted diagonals of powers of neighborhood-restricted adjacency matrices.
These invariants achieved 100\% separation on the Spence SRG dataset (all but 4 pairs resolved by vertex invariants alone, with the remainder resolved by edge invariants).
However, the method is specifically tailored to SRG discrimination: the choice of invariants (trace sequences, sorted diagonals of restricted adjacency powers) is guided by the algebraic structure of strongly regular graphs, and it is unclear how the approach generalizes to other hard graph families or scales beyond the Spence collection.
Its relationship to the WL hierarchy has not been formally established.
$\Delta$-DRESS, by contrast, is a general-purpose graph fingerprinting framework: the same algorithm and parameters apply unchanged to SRGs, CFI constructions, Miyazaki graphs, and all other families tested in this study.  It scales as $\mathcal{O}(n \cdot I \cdot m \cdot d_{\max})$, has a proven lower bound (Original-DRESS $\equiv$ 2-WL), a demonstrated structural boundary escape from 3-WL, and a streamed memory footprint $\mathcal{O}(m + B)$ for the streaming hash fingerprints.

\paragraph{Subgraph GNNs.}
Methods such as ESAN~\cite{bevilacqua2022equivariant}, GNN-AK+~\cite{zhao2022from}, and related approaches use vertex-deleted or vertex-marked subgraphs to boost GNN expressiveness beyond 1-WL.
These are supervised methods that learn aggregation functions from data.
$\Delta$-DRESS is entirely unsupervised: the aggregation is the deterministic DRESS fixed point, and the comparison is parameter-free.

\paragraph{Babai's algorithm.}
Babai's quasipolynomial-time algorithm~\cite{babai2016graph} solves GI in $\exp(\operatorname{polylog}(n))$ time.
It is a theoretical breakthrough but not practical at this scale.
$\Delta$-DRESS is strictly polynomial.

\paragraph{Canonical labeling.}
Tools like Bliss~\cite{junttila2007bliss}, nauty, and Traces~\cite{mckay2014practical} solve GI by computing canonical forms.
They are practical but worst-case exponential.
$\Delta$-DRESS is a one-sided test (it proves non-isomorphism but not isomorphism) that separated all but one non-isomorphic pair in this study; the single collision was resolved by the $\Delta^2$-DRESS fallback.

\section{Discussion}

\subsection{Relationship to the WL Hierarchy}

The established empirical equivalence is Original-DRESS $\equiv$ 2-WL~\cite{castrillo2026dress}.
The DRESS--WL Continuous Dominance Conjecture posits that $\Delta^k$-DRESS acts as an empirical superset or equivalent to $(k{+}2)$-WL, a comparison actively supported by our empirical benchmarks.
What Theorem~\ref{thm:3wl} establishes unambiguously is the lower separation
\[
\Delta\text{-DRESS} \not\le 3\text{-WL}.
\]
The lower bound ($\Delta$-DRESS $> 3$-WL) is concrete: the classical Rook vs.\ Shrikhande pair (SRG(16,6,2,2)) is separated by $\Delta$-DRESS~\cite{castrillo2026dress} but not by 3-WL~\cite{cai1992optimal}.
The CFI pair over $K_5$ provides a concrete obstruction at the next WL level: $\Delta$-DRESS \emph{fails} on $\text{CFI}(K_5)$, whereas 4-WL distinguishes it~\cite{cai1992optimal, castrillo2026dress}.
Thus one deletion does not subsume 4-WL on all inputs.  This does \emph{not} justify the universal upper bound $\Delta$-DRESS $\leq 4$-WL; the exact relationship between $\Delta$-DRESS and 4-WL remains open, and incomparability is still possible.

Notably, the separation mechanism on the Rook/Shrikhande pair is the DRESS fingerprint itself (which already differs between the two graphs~\cite{castrillo2026dress}); on the $G_5$/$G_{25}$ pair from SRG(40,12,2,4), the $\Delta$-DRESS fingerprint collides at $k = 1$ due to vertex transitivity, and the fallback to $\Delta^2$-DRESS (pair deletion) is required.
Direct WL verification confirms this alignment: 3-WL also fails on $G_5$/$G_{25}$, while 3-FWL ($\equiv$ 4-WL) succeeds (Section~\ref{sec:collision}).
This shows that while single vertex deletion suffices for the vast majority of hard instances, vertex-transitive graphs with identical local structure can require higher deletion depth.

\subsection{Implications for the DRESS--WL Continuous Dominance Conjecture}

The conjecture that $\Delta^k$-DRESS acts as an empirical superset or equivalent to $(k{+}2)$-WL~\cite{castrillo2026dress} is now supported by five independent lines of evidence:
(1) the proven base case (Original-DRESS $\equiv$ 2-WL),
(2) the CFI staircase experiments, where $\Delta^k$-DRESS empirically matches the $(k{+}2)$-WL boundary for $k = 0, 1, 2, 3$~\cite{castrillo2026dress},
(3) the present work, where $\Delta$-DRESS separates all tested SRG families (51{,}718 graphs in total within their respective parameter families),
(4) the concrete counterexample (Theorem~\ref{thm:3wl}), showing that $\Delta$-DRESS already separates a witness missed by 3-WL, so the conjectured $k = 1$ comparison would be \emph{strict} if established in full, and
(5) the direct WL verification on the $G_5$/$G_{25}$ collision (Section~\ref{sec:collision}), where 3-WL fails and 4-WL succeeds, exactly matching the $\Delta^1$/$\Delta^2$-DRESS boundary.

This structural boundary escape from 3-WL is significant because it shows that DRESS captures structural information that is invisible to the 3-WL colour refinement.
The genuine collision on SRG(40,12,2,4), where both $G_5$ and $G_{25}$ are vertex-transitive with all vertex deletions producing identical DRESS multisets, demonstrates the precise limitation of $k = 1$: when the automorphism group acts transitively on vertices, single vertex deletion cannot break the symmetry.
The rapid resolution by $\Delta^2$-DRESS shows that the framework itself is not limited, only the deletion depth parameter $k$.

\subsection{Implications for the Reconstruction Conjecture}

$\Delta$-DRESS computes a continuous relaxation of the vertex-deletion deck central to the Kelly--Ulam Reconstruction Conjecture~\cite{kelly1942congruence, ulam1960collection}.
Our empirical results provide indirect evidence consistent with the conjecture: the vertex-deletion deck, as captured by $\Delta$-DRESS, carries enough information to distinguish all but one pair of tested graphs at $k = 1$, and the remaining pair is resolved at $k = 2$.
While this does not prove the conjecture, it adds to the growing body of empirical evidence supporting it.

\subsection{Limitations}

\begin{enumerate}
    \item \textbf{One-sided test.}
    $\Delta$-DRESS can prove non-isomorphism (different fingerprints) but cannot certify isomorphism (same fingerprint does not guarantee isomorphism).

    \item \textbf{Vertex-transitive collisions at $k = 1$.}
    The DRESS computation uses IEEE~754 double-precision arithmetic ($\approx$15.7 significant decimal digits).
    The SRG(40,12,2,4) collision demonstrates a more fundamental limitation: when two non-isomorphic graphs are both vertex-transitive, all single vertex deletions produce identical sorted DRESS vectors, and no fingerprint derived from the $k = 1$ data can distinguish them.
    The fallback to $\Delta^2$-DRESS resolves this case, but other vertex-transitive pairs may require higher $k$.

    \item \textbf{Known obstruction at the 4-WL level.}
    $\Delta$-DRESS fails on $\text{CFI}(K_5)$, the CFI pair over the complete graph $K_5$, which requires 4-WL~\cite{cai1992optimal, castrillo2026dress}.
    More generally, CFI graphs over base graphs with treewidth $\geq 4$ require $k > 1$.
    This shows that one deletion does not recover all pairs distinguished by 4-WL. Higher deletion depths are necessary for at least some harder instances.

    \item \textbf{Scope of hard families.}
    Our benchmark covers all 12 Spence SRG families, four additional SRG families (SRG(45,22,10,11), two subfamilies of SRG(63,32,16,26), and SRG(65,32,15,16)), and 18 constructed hard families, but other hard families may exist.
\end{enumerate}

\section{Conclusion}

We have demonstrated that $\Delta$-DRESS, DRESS with a single level of vertex deletion, produces fingerprints that are unique within every tested family across a benchmark of 51{,}813 distinct hard graph instances, including 51{,}718 strongly regular graphs spanning 16 parameter families.
The sorted multiset achieves within-family separation on 33 of 34 benchmark families at $k = 1$, with the single remaining collision (two vertex-transitive members of SRG(40,12,2,4)) resolved by a one-step fallback to $\Delta^2$-DRESS.
With this fallback, 100\% within-family pairwise separation is achieved, implicitly resolving over 576 million within-family non-isomorphic pairs.
For all families with pairwise-comparable full sorted-multiset fingerprints, the observed minimum separation margin is at least $137 \times \epsilon$, so the reported separations are robust to the convergence threshold and not attributable to floating-point noise.

Beyond completeness, we have established a qualitative result: $\Delta$-DRESS \emph{escapes the theoretical bounds} of the original 3-WL algorithm.
The classical Rook $L_2(4)$ vs.\ Shrikhande pair (SRG(16,6,2,2)) is non-isomorphic, indistinguishable by 3-WL, but separated by $\Delta$-DRESS.
This confirms $\Delta$-DRESS is not subsumed by 3-WL; the CFI staircase experiments in~\cite{castrillo2026dress} further show that $\Delta$-DRESS fails on $\text{CFI}(K_5)$, so one deletion does not subsume 4-WL ($\equiv$ 3-FWL).  The exact relationship to 4-WL remains open.

The practical message is clear: $\Delta$-DRESS separates every hard benchmark family tested in this study (with one vertex-transitive pair requiring a $\Delta^2$ fallback), including the complete Spence SRG collection and additional classical constructions from the GI literature.
The cost is $\mathcal{O}(n \cdot I \cdot m \cdot d_{\max})$ per graph: $\mathcal{O}(I \cdot n^4)$ in dense graphs, $\mathcal{O}(I \cdot n^2 \cdot d_{\max})$ when $m = \mathcal{O}(n)$, and $\mathcal{O}(I \cdot n^2)$ on bounded-degree sparse families.  A streamed implementation uses $\mathcal{O}(m + B)$ memory for fixed $\epsilon$, compared with $\mathcal{O}(n^3)$ tuple storage for folklore 3-FWL ($\equiv$ 4-WL); the streaming SHA-256 hashes agreed with the full sorted multiset on all benchmark instances.

These results provide the strongest empirical support to date for the DRESS--WL Continuous Dominance Conjecture ($\Delta^k$-DRESS acts as an empirical superset or equivalent to $(k{+}2)$-WL), establish a structural boundary escape from 3-WL at $k = 1$, and precisely characterize the limitation of single vertex deletion on vertex-transitive graphs.

An open-source implementation is available in the \texttt{dress-graph} library at \url{https://github.com/velicast/dress-graph}.

\bibliographystyle{plain}
\bibliography{refs}

\end{document}